\documentclass[10pt]{article}

\usepackage{lineno}

\usepackage{amsmath}
\usepackage{amssymb}

\usepackage{graphicx}

\usepackage{cite}

\usepackage{color}

\usepackage[labelfont=bf,labelsep=period,justification=raggedright]{caption}

\makeatletter
\renewcommand{\@biblabel}[1]{\quad#1.}
\makeatother

\date{}

\usepackage[lofdepth,lotdepth]{subfig}

\begin{document}

%Change title
%Check commented bibliography and, possibly, iclude it
%Check recent literature, especially PLOS and Arkin
%Include some discussion of the definition of memory as rate-independent limit with references to hysteresis literature

% Title must be 150 characters or less
\begin{flushleft}
{\Large
\textbf{Hysteresis Can Grant Fitness in Stochastically Varying Environment}
}
% Insert Author names, affiliations and corresponding author email.
\\
Gary Friedman$^{1}$,
Stephen McCarthy$^{2}$,
Dmitrii Rachinskii$^{2,3,\ast}$
\\
\bf{1} Department of Electrical and Computer Engineering, Drexel University, Philadelphia, PA, USA
\\
\bf{2} Department of Applied Mathematics, University College Cork, Cork, Ireland
\\
\bf{3} Department of Mathematical Sciences, University of Texas at Dallas, Richardson, TX, USA
\\
$\ast$ E-mail: dmitry.rachinskiy@utdallas.edu
\end{flushleft}

% Please keep the abstract between 250 and 300 words
\section*{Abstract}
Hysteresis and bet-hedging (random choice of phenotypes) are two different observations typically linked with multiplicity of phenotypes in biological systems. Hysteresis can be viewed as form of the system's persistent memory of past environmental conditions, while bet-hedging is a diversification strategy not necessarily associated with any memory. It has been shown that bet-hedging can increase population growth when phenotype adjusts its switching probability in response to environmental inputs. Although memory and hysteresis have been used to generate environment dependent phenotype switching probabilities, their exact connection to bet-hedging have remained unclear. In this work, using a simple model that takes into account phenotype switching as well as lag phase in the population growth occurring after the phenotype switching, it is shown that memory and hysteresis emerge naturally and are firmly linked to bet-hedging as organisms attempt to optimize their population growth rate. The optimal ``magnitude'' of hysteresis is explained to be associated with stochastic resonance where the characteristic time between subsequent phenotype switching events is linked to the lag phase delay. Furthermore, hysteretic switching strategy is shown not to confer any additional population growth advantage if the environment varies periodically in a deterministic fashion. This suggests that, while bet-hedging may evolve under some conditions even in deterministic environments, memory and hysteresis is probably the result of environmental uncertainty in the presence of lag phase in the switching of phenotypes.

\section*{Introduction}
\begin{figure}[h]
\begin{centering}
\includegraphics[width=0.45\columnwidth]{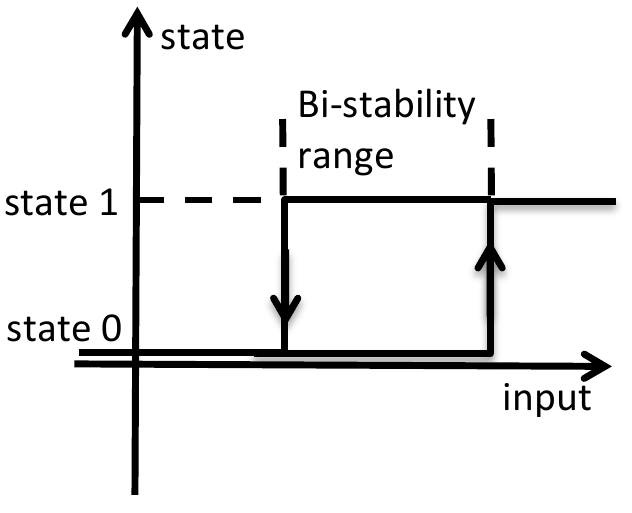}\qquad \includegraphics[width=0.48\columnwidth]{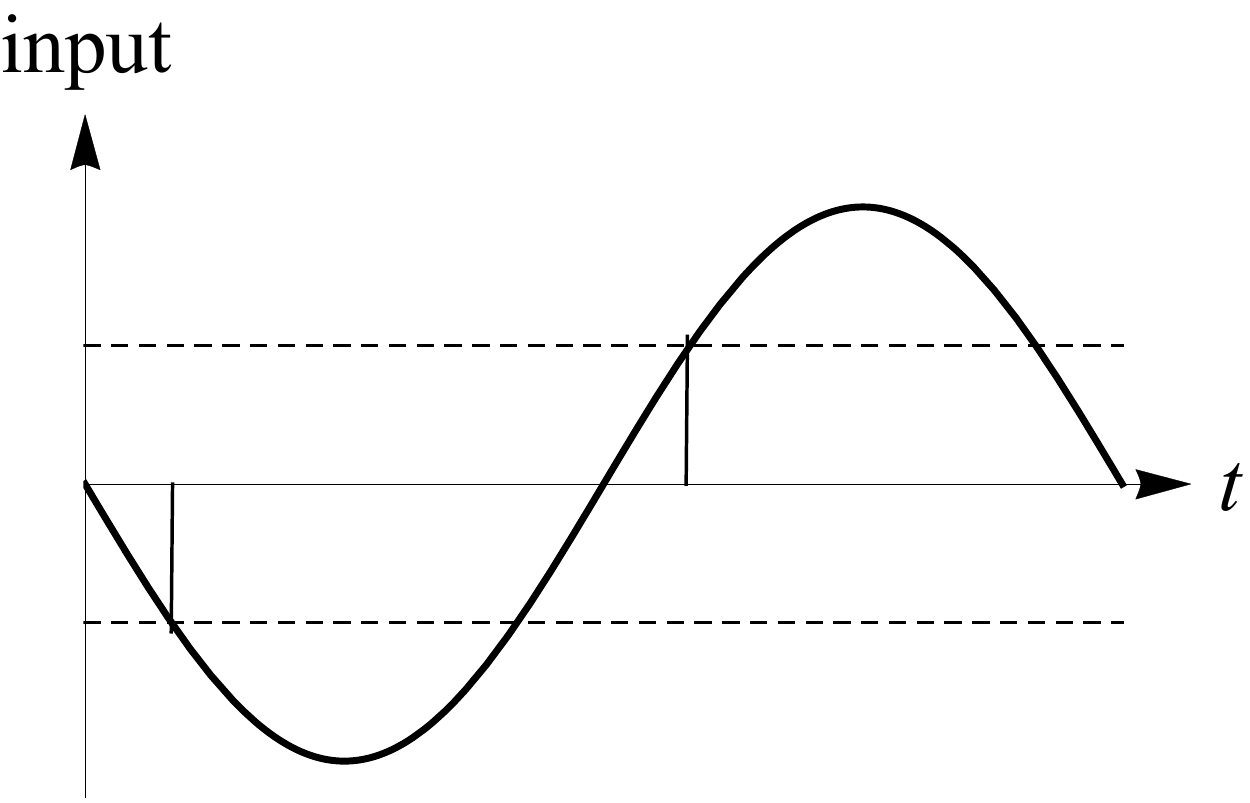}
\end{centering}
\centerline{(a) \hskip7.5truecm (b)
}
\caption{Illustration of bi-stable relay hysteresis. When the state is 1, the relay switches to state 0 at the lower input threshold,
while starting from state 0 the relay switches to state 1 at an upper threshold. Thus, as long as the current input is within the
bi-stability range (i.e. between the thresholds), the relay remembers whether the input has entered this range from below or from above. Panel (a) shows the input-state diagram.
Panel (b) presents an example of input graph. The vertical interval between the horizontal dashed lines on panel (b) corresponds
to the horizontal bi-stability interval on panel (a). The intersection of the lower (upper) dashed line with the input graph on (b)
defines a moment of switching from state 1 to 0 (from 0 to 1).}
\label{relay}
\end{figure}

The term ``hysteresis'' seems to have been coined by James Alfred Ewing \cite{ewing} in connection with the ability of some magnetic materials to retain their magnetization state long after the magnetizing magnetic field has been removed. Today it is used much more broadly to refer to any memory based relationship between an input and state of a system that does not depend on the rate at which the input varies in time \cite{SH}. The most basic, yet non-trivial hysteresis is exemplified by a bi-stable relay operator illustrated in Figure \ref{relay}.  This type of hysteresis determines the current output on the basis of the current input and the most recent previous output.

Although the bi-stable relay or other types of hysteresis are idealized descriptions of memory, they are useful because they are dramatically different from the more common forms of dynamical system memory. While dynamic memory is associated with certain characteristic times and can be obtained even in simple linear systems, hysteretic systems remember input features such as input extrema and this memory can persist for an unspecified time. In contrast to dynamic memory, hysteresis relies on the ability of a system to remain in one of at least two (and usually many more) stable states for any given input and this is its primary connection with multi-stability in biological systems.

Potential importance of multi-stability in living systems has been articulated first by Max Delbr\"uck \cite{delbr}, who associated different stationary states with epigenetic differences in clonal populations such as those arising in the process of cell differentiation. The simplest classical example is probably bi-stable behavior of {\em lac-operon} in {\em E.~coli}. {\em Lac-operon} is a collection of genes associated with transport and metabolism of lactose in the bacterium. Expression of these genes can be turned on by certain small molecules that have been called inducers. Novick and Weiner \cite{f0} as well as of Cohn and Horibata \cite{cohn1,cohn2,cohn3}, relying on prior work of others \cite{monod,benzer,Spiegelman} effectively demonstrated that two phenotypes each associated with ``on'' and ``off'' state of {\em lac-operon} expression can be obtained from the same culture of genetically identical bacteria. The fraction of each corresponding sub-population depended on the history of exposure to the inducer. As illustrated in Figure \ref{relay}, the {\em lac-operon} state was induced (switched on) when the extracellular inducer concentration (input) exceeded an upper threshold. The operon was switched off when the inducer concentration fell below the lower threshold. Both phenotypes remain stable through multiple generations of the bacterial culture after the extracellular concentration of the inducer is reduced to lower levels, but not removed completely. Novick and Weiner did not use the term ``hysteresis'' to describe their observations, but effectively that is what it was.

These early findings on the hysteresis of the {\em lac-operon} enzymes were consistent with earlier findings on stability of other enzymes in yeast \cite{winge}.  Recent experiments using molecular biology methods (such as those incorporating green fluorescent protein expression under the {\em lac-operon} promoter) permitted to confirm and further study the region of bi-stability of the {\em lac-operon} even when multiple input variables (TMG that acts as the inducer and glucose, for example) are used to switch the {\em lac-operon} genes on and off
\cite{oud}. Multi-stable gene expression and hysteresis have been well-documented in many natural as well as artificially constructed biological systems \cite{wanga,lai,grazi,f3,bac4,smits,maamar,ark,c10}.

Several things are probably worth noting when it comes to hysteresis and multi-stability in biological systems. First is that they are consistently associated with difference in the reproductive rates of different phenotypes in different environments. This was noted in the experiments reported by Novick and Weiner, for example, and suggests association of phenotype hysteresis and multi-stability with strategies for increasing population growth in varying environments.

Second relates to the very essence of hysteresis - rate-independence. In reality rate-independence holds only over a certain intermediate time scale and does not hold for very quickly or very slowly changing environments. If the environmental stimulus, such as an extracellular inducer concentration for the {\em lac-operon}, is varied too rapidly, the cell simply may not have sufficient time to complete its switching from one phenotype to another. Internal switching dynamics will invariably require some characteristic time. In some cases, this time may be associated with molecular transport across the cell membrane. In other cases, this time may be required to produce sufficient amount of some intracellular factor. In micro-organisms this internal dynamics is associated with a well-known delay time that exhibits itself through a lag phase that often occurs when some environmental input is changed quickly.

At the other time scale extreme, if the inducer concentration changes slowly or is maintained at a constant value within the region of bi-stability over a long time period typically involving multiple generations, random fluctuations would cause the population to forget its history eventually \cite{mcadams,matheson}. This is typically evidenced by a stable history independent bi-modal (and more generally multi-modal) distribution of phenotype features \cite{siegele, kaern, elowitz,isaacs,bec}. Such memory erasing random fluctuations are also the key challenge in magnetic information storage technology. The probability of random switching from a given phenotype typically increases as the average value of an environmental input moves closer to the corresponding switching threshold. Therefore, adjustment of the switching threshold values and tuning of the switching probabilities are related and hysteresis has been viewed as mechanism by which cells can implement and adapt a diversification strategy of environment dependent random phenotype switching.

The idea of bet-hedging through diversification of phenotypes has been discussed in different biological contexts as an adaptation strategy to time-varying environments in many publications often without reference to any other specific mechanism by which it can be implemented \cite{bac2}. The main premise behind it is that genetically identical organisms can grow faster by choosing some fraction of their population to be in a phenotype less favored in the current environment, but prepared for potentially different future environment \cite{beau}. This can be viewed as an inherently pessimistic strategy of survival: organisms switch to the less favorable phenotype in anticipation of the worst. Typical models of phenotype diversification describe population growth by a set of first-order coupled differential equations where the growth of each phenotype is proportional to populations of each phenotype. Coefficients of proportionality in this system of equations can be interpreted as growth rates and phenotype switching rates.

A simple model for the system with two phenotypes proposed in \cite{bac1} describes this as follows. Let the populations of the two phenotypes be $x (t)$ and $y (t)$ and let some environmental input $E(t)$ be characterized by a binary variable taking on values of $0$ or $1$. The growth of the population is described by two differential equations:
\begin{equation}\label{e1}
\begin{array}{l}
x '=\gamma_1 (E) x -k_1 (E) x +k_2(E) y \\
y '=\gamma_2 (E) y -k_2 (E) y +k_1 (E) x
\end{array}
\end{equation}
where ``prime'' denotes time derivative, $k_1 (E)$ and $k_2 (E)$ are environment dependent rates at which organisms switch from phenotypes $x $ and $y$, respectively, while $\gamma_1 (E)$ and $\gamma_2 (E)$ are the corresponding phenotype growth rates. This type of model is appropriate for a non-competitive process of population growth where local resources are not limited. The long-term growth of the population in this model was evaluated by the Lyapunov's exponent:
\begin{equation}\label{e2}
\lambda= \lim_{t\to\infty}⁡ \left\{\frac{1}{t} \int_0^t\frac{\gamma_1 (E(\tau)) x (\tau)+\gamma_2 (E(\tau)) y (\tau)}{x (\tau)+y (\tau)} d\tau \right\}
\end{equation}

The above model demonstrated that, when the environment varies with some period and different phenotypes are favored in different environments (i.e.~$\gamma_1 (E=0)>\gamma_2 (E=0)$, but $\gamma_2 (E=1)>\gamma_1 (E=1))$, the long-term growth is maximized when the switching rate out of the favored phenotype is non-zero as long as transition rates are bounded to be below some value. Similar behavior holds in uncertain environments (Poisson distribution of environmental durations was considered). Experimental work by some of the same authors provided a substantial support for these findings \cite{bac5,bac6,kauf}.  Other models considered more than one possible stable state of the phenotype and of the environment taking the phenotype switching rates to be independent of the environment in the random switching regime and concluding that phenotype switching probabilities should be matched to environment switching probabilities \cite{bac2, bac5}. However, it seems that the issue of hysteresis always remained uncoupled from the random phenotype choice strategy. It is demonstrated below that these two phenomena (hysteresis and diversification) can be coupled when lag phase delay in population growth is taken into account.

\section{Model}

Two modifications will be introduced in this work to refine model (1). Firstly, the environmental variable will not be restricted to being a binary. Instead, it will be permitted to vary continuously. Secondly, the model will incorporate explicitly the possibility of a lag phase between switching events by introducing additional populations $w$ and $z$ corresponding to organisms waiting to switch into phenotypes $x$ and $y$, respectively. Incorporating the lag phase is important from several different points of view. On the one hand, it effectively introduces a realistic opportunity cost associated with any phenotype switching event since the organism lose the opportunity to reproduce while waiting in the lag phase. On the other, it reflects a characteristic time inherent to cellular dynamic processes linked to phenotype switching. With these modifications the growth model becomes:
\begin{equation}\label{e3}
\begin{array}{l}
x'=\gamma_1 (E)x-k_1 (E)x+\delta w     \\
y'=\gamma_2 (E)y-k_2 (E)y+\delta z    \\
z'=k_1 (E)x-\delta z           \\
w'=k_2 (E)y-\delta w
\end{array}
\end{equation}
where the lag phase characteristic time is $1/\delta$.

Different functional dependences of the coefficients in model (3) can be considered. Continuous functions (piecewise linear sigmoidal) are used here to model dependence of growth coefficients on the environment:
\begin{equation}\label{e5}
\gamma_1(E)=\left\{
\begin{array}{cl}\gamma+\sigma,  & E\le 0\\
   -\sigma(E-0.5)+0.5 \sigma+\gamma, & 0<E<1;\\
   \gamma, & E\ge 1
   \end{array}\right.\qquad
                  \gamma_2(E)=2\gamma+\sigma-\gamma_1 (E)
                  \end{equation}
where $\gamma$ is the minimum growth rate possible and $\gamma+\sigma$ is the maximum possible growth rate. Effectively, $\sigma$ is the maximal growth rate advantage of the alternate phenotype.

Although the above dependence is employed here primarily to illustrate key features of the model, it can be viewed as reasonable because 1) the growth rates of both phenotypes can be the same at some value of the environmental variable (which is set to $E=0.5$ here) and 2) only a partial favoring of one phenotype over the other is possible when the environmental variable deviates from $E=0.5$ by a small amount. On the other hand, at large deviations of the environmental variable from its average, the difference in favoring one phenotype over another is bounded by some value $\sigma$.

Similarly to the previously considered model (1), the dependence of the switching rates $k_1 (E)$, and $k_2 (E)$ on the environmental variable will be described by step functions. However, in contrast to the previous model, the thresholds for these steps will not be required to be the same for the two phenotypes. We use a parameter $\alpha$ to specify the thresholds:
\begin{equation}\label{e4}
k_1(E)=\left\{\begin{array}{l}
k_u,\ E\le 0.5+\alpha\\
k_f,\ E>0.5+\alpha
\end{array}\right.\qquad
k_2(E)=\left\{\begin{array}{l}
k_f,\ E\le 0.5-\alpha\\
k_u,\ E>0.5-\alpha
\end{array}\right.
\end{equation}
where %$k_u$ and $k_f$ are switching rates such that
%\begin{equation}\label{5'}
%k_u+k_f=k;
%\end{equation}
 $k_u$ is the rate at which bacteria switch from favored to unfavored phenotype and $k_f>k_u$ is the rate of switching from unfavored to favored phenotype. When $\alpha=0$, the change of the transition rate from one phenotype to another coincides with the change of the phenotype growth status from favored to unfavored or vice versa. This change of transition rates can be characterized as ``realistic" strategy. Positive $\alpha$ implies that there is an interval of the environmental input $ 0.5-\alpha< E< 0.5+\alpha$ over which both phenotypes have low transition rates $k_u$. In this case one phenotype retains its low transition rate even after its growth status has changed to unfavored, while the other reduces its transition rate even before its growth status changes to favored. For this reason the strategy corresponding to positive $\alpha$ can be characterized as optimistic. Negative $\alpha$ means that both phenotypes have high transition rate $k_f$ over the interval $0.5+\alpha< E< 0.5-\alpha$ which corresponds to a pessimistic strategy, although somewhat different from bet-hedging.
	
Variation of the environment will be modeled here in two distinct ways: by a random process and by a periodic symmetric ($E(t)=1-E(t-T/2)$) function. One of the random processes employed here is the Ornstein-Uhlenbeck (OU) process that describes diffusion-like motion in a one-dimensional parabolic potential centered at the point $E=0.5$
where the growth rates of the phenotypes are equal:
\begin{equation}\label{e7}
dE=-a(E-0.5 )dt+dW_t
\end{equation}
where $a$ is a stiffness parameter associated with the parabolic potential well and $dW_t$ is the derivative of the Weiner process (white noise) creating stochastic fluctuations around
the point $E=0.5$. Average time $\tau_E$ of passage of the interval $0.5-|\alpha|\le E\le 0.5+|\alpha|$ by the OU process can be viewed as a certain characteristic time of this process. A modification of the OU process that corresponds to a double-well, rather than a parabolic potential will also be considered:
\begin{equation}\label{e8}
dE=(a(E-0.5)-(E-0.5 )^3 )dt+dW_t
\end{equation}
Deterministic periodic environmental variation will be taken as sinusoidal having a half-period $\tau_E$.

The effect of the threshold $\alpha$ on the population growth will be investigated along with the effects of parameters $k_{u,f}$ and $\delta$ using the Lyapunov's exponent to represent the asymptotic growth:
\begin{equation}\label{e9}
\lambda=\lim_{t\to\infty}\left\{ \frac{1}{t} \int_0^t \frac{\gamma_1 (E(\tau))x(\tau)+\gamma_2 (E(\tau))y(\tau)}{x(\tau)+y(\tau)+w(\tau)+z(\tau)} d\tau\right\}
\end{equation}

\section*{Results}

\begin{figure}[h]
\centerline{
\includegraphics[width=.46\textwidth]{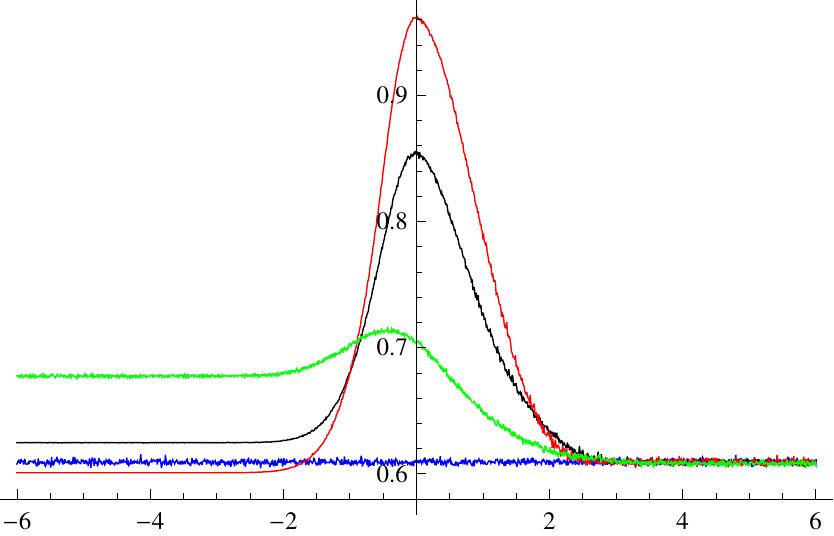} \quad \includegraphics[width=.46\textwidth]{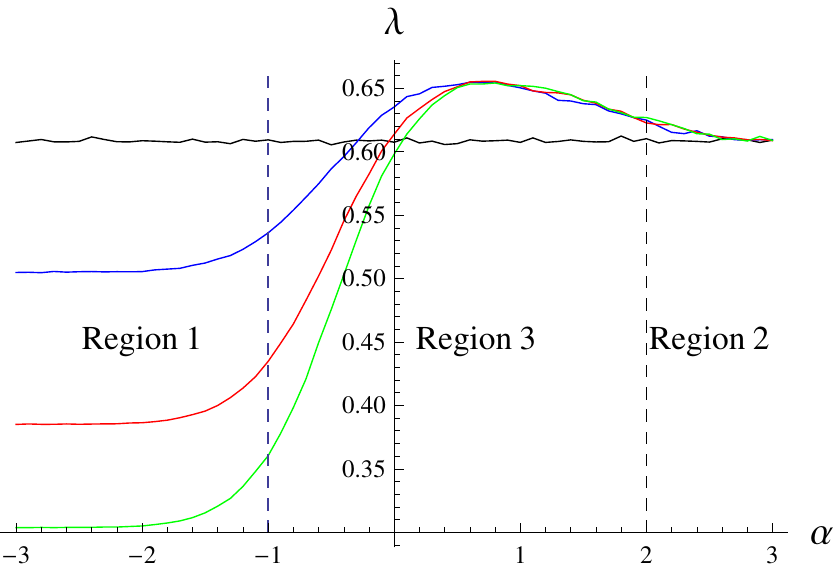}}
\centerline{(a) \hskip7truecm (b)
}
\caption{Plots of the Lyapunov's exponent $\lambda$ obtained by averaging 20 simulations of (a) system~\eqref{e1} and (b)
system~\eqref{e3} with OU environmental input \eqref{e7} for different $\alpha$ values. (a) The blue, green, black, and red lines are for $k_f=0, 0.3, 2.6, 100$, respectively. (b)
 The black, blue, red and green lines are for $k_f=0, 0.5, 1, 1.5$, respectively.
Other parameters are $k_u=0$, $\sigma=1$, $a=1$, $\gamma=0.1$,  $\delta=1$.}
  \label{fig2}
\end{figure}

 %Axes captions
% Length of the trajectory (time interval)?
% Initial values

The results are presented for zero transition rate $k_u=0$ from favored to unfavored phenotype.

When the lag time is close to zero (large $\delta$), the results obtained from models (\ref{e3}) and (\ref{e1}) are similar.
Figure~\ref{fig2}(a) presents the dependence of the average growth rate $\lambda$ on $\alpha$ for the system with zero lag phase, which is driven by OU environmental process (\ref{e7}).
The blue horizontal line corresponds to $k_f=0$, the case of no flow between phenotypes. Here the Lyapunov's exponent $\lambda$ is close to the arithmetic mean $\gamma+\sigma/2$ of the saturated highest and lowest growth rates.
%The variability of the line is a result of the stochastic nature of the input $E$ to the system and should be eliminated by the inclusion of more simulations
%(or, increasing the time interval of a simulation).
The other three curves corresponding to positive transition rate $k_f$ from unfavored to favored phenotype demonstrate a clear maximum, which
increases with $k_f$.
These curves converge to the blue curve as $\alpha$ increases, the reason being that for large positive $\alpha$ the switching threshold values are so high that the environmental input reaches them rarely, hence little switching occurs and the system behaves almost as in the no flow case $k_f=0$.
A different behaviour is observed
for large in absolute value negative $\alpha$. In this limit, the environmental input is unable to pass the threshold values so as to turn off the switching rate, hence each phenotype is constantly transitioning into the other with the effect that the populations
%. Effectively, in this case, the transition rates in \eqref{e1} are positive and constant all the time, $k_{1,2}(E)=k_f$.
%Therefore, a large in absolute value negative $\alpha$ creates a constant back flow from an unfavoured to the favoured phenotype,
% which has the same rate as the opposite flow, resulting
are permanently mixing at constant rate $k_f$. Figure~\ref{fig2}(a) shows that the Lyapunov's exponent $\lambda$
%is higher than that obtained for no transitions between states, but as $k_f$ increases, $\lambda$ becomes lower than that obtained for no transitions.
first increases and then decreases with increasing $k_f$ for large in absolute value negative $\alpha$.
%This behaviour is consistent with the results of \cite{bac1} where it was shown (for an environment stochastically switching between two states) that, for a bounded range of values of the %transition rate from an unfavoured to the favoured phenotype, the average growth rate could be maximised by having a certain rate of transition from a favoured into the unfavoured %phenotype (back flow).
%
%For low transition rates $k_f$, the peak in the curves occurs for predictive switching ($\alpha<0$), that is the system begins switching of phenotypes before the environment changes which %phenotype it favours. As the transition rate increases, there is a shift in the location of the peak to $\alpha=0$ corresponding to the responsive strategy where the system undergoes %switching at the same moment as the environment starts to favour one phenotype over the other.
%
Furthermore, when the switching rate $k_f$ is relatively low, the maximal growth rate $\lambda$ is achieved by the
bet-hedging (pessimistic) strategy corresponding to a negative value of $\alpha$, while for
  larger $k_f$ the optimal asymptotic growth occurs for the realistic strategy corresponding to $\alpha=0$.
  This behavior is consistent with the results of  \cite{bac1} where $\alpha$ was always zero,
  but a positive transition rate $k_u$ played the role of an (alternative) bet-hedging mechanism. Increasing $k_u$ with simultaneously setting a negative $\alpha$
  help increase the growth rate in the present model in case of a relatively low switching rate $k_f$
  (not shown in the figure).

Figure \ref{fig2}(b) shows that optimal asymptotic growth occurs at positive values of $\alpha$ when the lag time $1/\delta$ becomes non-zero. Interestingly, the optimal value of $\alpha$ is nearly independent of the maximal switching rates as long as the lag delay is the same.

%A comparison of Figs.~\ref{fig2}(a) and (b), which present results of numerical integration of systems \eqref{e1} and \eqref{e3}, respectively, with the input
%\eqref{eq::e}, shows that the introduction of the switching cost in the form of temporary inhibition of the reproduction in the model has
%a substantial effect on the optimal switching strategy. The average growth rate $\lambda$ for models \eqref{e1} and \eqref{e3} is defined by the same formula \eqref{eq::averagegrowthrate}.
The plot in Figure~\ref{fig2}(b) is divided into three regions, $\alpha\leq-1$ (region 1), $\alpha\geq2$ (region 2), and $-1<\alpha<2$ (region 3). The horizontal line corresponding to $k_f=0$, the case where there is no transitions between phenotypes, is the same as in Figure~\ref{fig2}(a).
In region 1, the Lyapunov's exponent $\lambda$ rapidly decreases with $k_f$.
The reason is that the environment is between the thresholds most of the time for this region, $0.5-\alpha<E(t)<0.5+\alpha$,
hence the majority of bacteria are nearly always in a transition state due to the pessimistic strategy (negative $\alpha$).
%For red, blue and green lines, the absolute value of $\alpha$ is sufficiently large so that the environmental input $E$ rarely reaches the thresholds of $k_1$ and $k_2$. Since $\alpha$ is %negative and the thresholds of the functions $k_1$ and $k_2$ are rarely reached, the majority of bacteria are nearly always in a transition state.
The higher the transition rate $k_f$, the higher the fraction of the total population that is stuck in the groups $z$ and $w$ that do not contribute to the growth of the system, hence lower $\lambda$.
In region 2, the value of $\alpha$ is also sufficient large so that the environmental input mostly remains within the bi-stability interval. As $k_u=0$,
              both rates $k_{1,2}(E)$ are nearly always zero due to the optimistic strategy corresponding to positive $\alpha$ in this region, hence the majority of bacteria are in the non-transition states $x,y$ and the plots of $\lambda$ for all $k_f$ tend to the horizontal plot obtained for $k_f=0$ as $\alpha$ increases.
 Central Region 3 is the most interesting as each plot $\lambda(\alpha)$ corresponding to a non-zero value of $k_f$ achieves a distinct global maximum at some positive
 value of $\alpha$, that is for the optimistic switching strategy.
 %This means that the population achieves the maximum average growth rate by adopting the delayed switching strategy. That is, bacteria delay switching phenotypes until the environment has a certain level of favouring of the other phenotype.
%This delayed switching behaviour is beneficial to the system as a whole as it eliminates any switching that may result
%from small fluctuations of the environment around the value $E_T$ if $\alpha$ is zero or small.
%The drop in $\lambda$ caused by such switching due to the inhibition of the growth rate in a transition state ($z$ or $w$)
%is bigger then the drop in $\lambda$ caused by a small and short drop in the growth rate of a reproductive state ($x$ or $y$) when switching is eliminated
%by a larger $\alpha$.

%As the peak of the curve is shifted from zero $\alpha$ in Fig.~\ref{fig2}(a) to the region of positive $\alpha$ in Fig.~\ref{fig2}(b), we see that,  when growth is inhibited by switching, %the optimal behaviour changes from the memoryless responsive strategy to the delayed strategy.

\begin{figure}[h]
\centerline{
\includegraphics[width=0.46\textwidth]{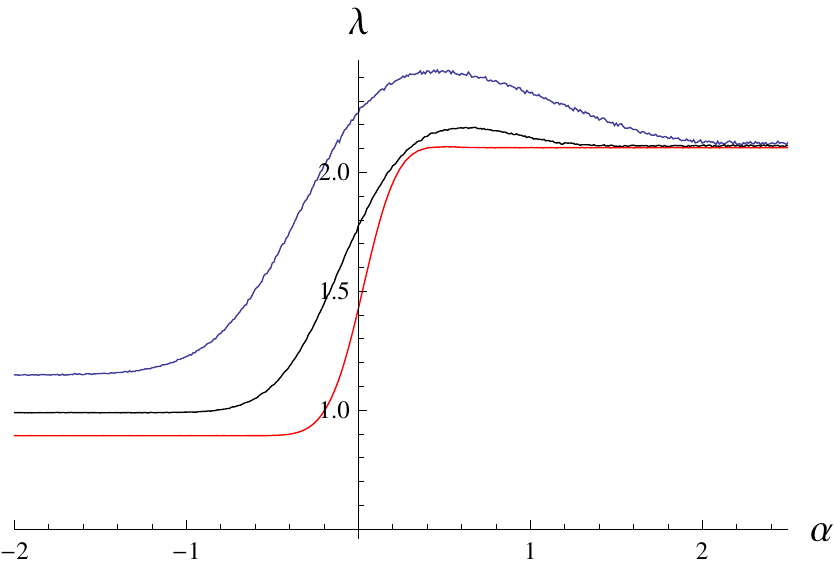}
\label{subfig::steep}
\quad
\includegraphics[width=0.46\textwidth]{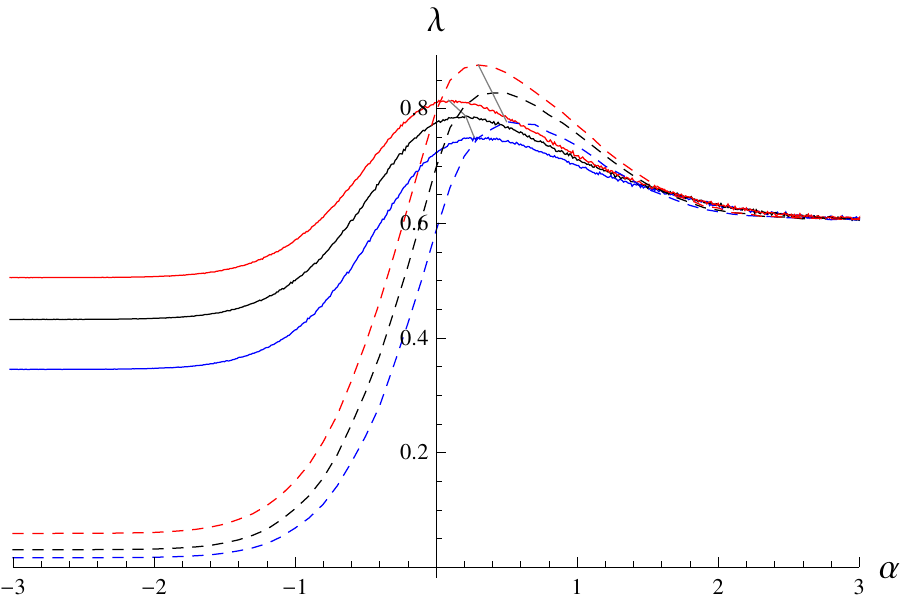}
}
\centerline{(a) \hskip7truecm (b)
}
\caption{\label{fig::parametervariation1}The effect of changes in the parameters $a$ and $\delta$ of model \eqref{e3} on the Lyapunov's exponent $\lambda$
and the optimal value of the threshold parameter $\alpha$, which maximizes $\lambda$.
(a) The effect of varying the stiffness of the potential well $a$. The blue, back, and red lines correspond to $a=2, 5, 21$, respectively. Other parameters are $k_u=0$, $\sigma=4$, $k_f=2.6$, $\delta=1$ and $\gamma=0.1$. (b) The effect of altering the average lag time $1/\delta$. The blue, black, and red lines correspond to $\delta=2.8, 5.3, 10.3$, respectively.
Solid lines are plotted for $k_f=2.6$, a relatively slow transition rate; dashed lines are plotted for $k_f=100$, a high transition rate.
Other parameters are $k_u=0$, $\sigma=1$, $a=1$ and $\gamma=0.1$.}
\label{fig3}
\end{figure}

Now, we consider how system \eqref{e3} responds to variations of parameters.
%the model parameter $\delta$, which controls how long bacteria stay in the non-reproductive delay groups;
%the difference of the growth rates of a favoured and unfavored phenotypes $\sigma$; and, the stiffness of the potential well $a$, which controls fluctuations of the environment.
%The results are presented in Figs.~\ref{fig::parametervariation1} and \ref{fig::parametervariation2}.
Figure \ref{fig3} illustrates dependence of positive value of $\alpha$ needed to obtain maximum asymptotic growth on the  ``stiffness'' parameter of the OU process and on the lag time.
Examining the plots showing the dependence $\lambda(\alpha)$ in Figure~\ref{fig3}(a) for several values of the stiffness of the potential well of the environmental input, $a$, we see that as the well becomes steeper and the environmental input is forced to spend more time around the point $E=0.5$ of equal favoring of the phenotypes, the value of the peak in the Lyapunov's
exponent $\lambda$ decreases. When the well becomes sufficiently steep, the peak is lost and the Lyapunov's exponent converges to the value $\gamma+0.5\sigma$ of the
average growth rate of the system with no transitions between phenotypes.

Figure \ref{fig3}(b) shows that, as the lag time $1/\delta$ increases, the optimal positive value of $\alpha$ which grants the maximal Lyapunov exponent also increases.
%Fig.~\ref{fig3}(b) shows the effect that variations in the length of the delay characteristic times, $1/\delta$, have on the graphs $\lambda(\alpha)$. Increasing the value of $\delta$ is %equivalent to shortening the time a bacterium in transition spends in the non-reproductive state (groups $z$ and $w$). We see that for increasing values of $\delta$ the peak
%of the average growth rate moves left towards $\alpha=0$.
This trend is in agreement with Figure \ref{fig2}(a) presenting the limit case of zero lag time
where the corresponding optimal $\alpha$ is zero or negative.
%\textcolor{red}{Mechanism of the dependence of the optimal $\alpha$ on the lag time and stiffness is related to the dependence of the average time required for the OU process to pass %through the bi-stability interval, see Figure \ref{fig4}(a). There is a direct relationship between the bi-stability interval exit time required to optimize the asymptotic growth and the %lag time. -- either add figure or delete this phrase.}

\begin{figure}[h]
\centerline{\includegraphics[width=0.46\textwidth]{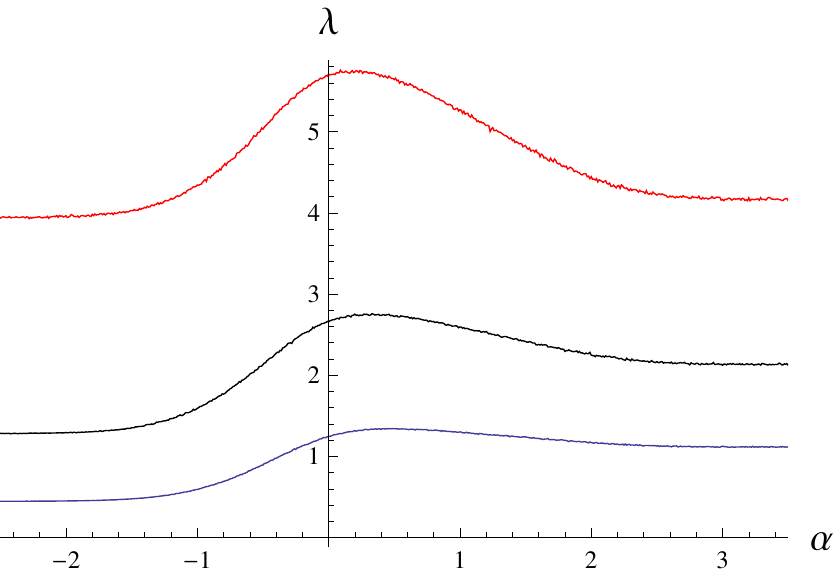}}
\caption{Results of modifying the difference of the growth rates of the fully favored and fully unfavored phenotypes, $\sigma$. The blue, black, and red lines correspond to $\sigma=2,4,8$, respectively. Other parameters are $k_u=0$, $k_f=2.6$, $\delta=1$, $a=1$ and $\gamma=0.1$.}
\label{fig4}
\end{figure}

In Figure~\ref{fig4}, we vary the parameter $\sigma$, which controls the benefit to the growth rate that bacteria in a favored phenotype gain
over bacteria in the unfavored phenotype. Increasing the value of $\sigma$ has an effect similar to that of shortening the lag time by increasing $\delta$,
cf. Figure \ref{fig3}(b). This result can be understood if we consider $\sigma$ as a penalty for being in the wrong phenotype when the environment changes. When the penalty becomes too high it is no longer worth delaying changing phenotype and becomes better to change with the environment using the realistic switching strategy, that is setting $\alpha=0$.

%Alterations to the parameter $\gamma$, the growth rate of the unfavored phenotype, simply cause a translation of the plots up the $\lambda$ axis; that is, $\gamma$ is an additive constant %to the average growth rate.

\begin{figure}[ht]
\centerline{\includegraphics[width=0.46\textwidth]{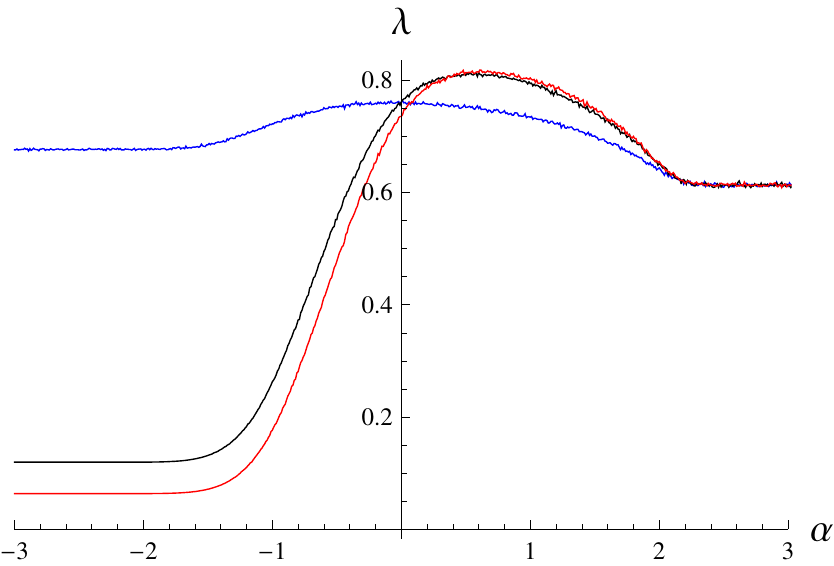}
\quad
\includegraphics[width=0.46\textwidth]{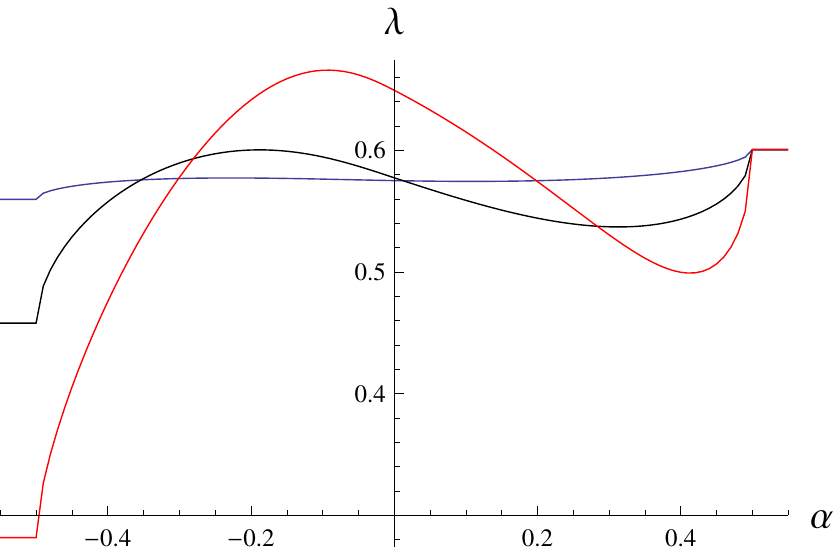}
}
\centerline{(a) \hskip7truecm (b)
}
\caption{\label{fig::AltEnviron}Dependence of the Lyapunov's exponent $\lambda$ on the parameter $\alpha$ for alternative environmental inputs to model \eqref{e3}.
(a) Results for the stochastic input \eqref{e8}. The blue, black, and red lines correspond to $k_f=0.2, 4.7,9.2$, respectively.
(b) Results for the periodic environmental input $E(t)=0.5(1+\sin t)$. The blue, black, and red lines are for $k_f=0.45, 0.7, 1.7$, respectively.
 Other parameters are the same as in Figure \ref{fig2}.}
\label{fig5}
\end{figure}

Finally, we test system \eqref{e3} with environmental inputs different from the OU process.
Figure~\ref{fig5}(a) presents data obtained for input \eqref{e8} generated by the diffusion process in a double well potential. When the transitions rate $k_f$ is low, the Lyapunov's exponent is maximized by $\alpha=0$. However, as the transition rate $k_f$ increases, the peak in the Lyapunov's exponent profile $\lambda(\alpha)$ shifts to the region of positive $\alpha$. That is, as in case of environmental input \eqref{e7} (see Figure~\ref{fig2}(b)), the optimistic strategy grants more fitness to the population than the realistic strategy
for non-zero lag times.
Plots in Figure~\ref{fig5}(b) were obtained for a periodic input, which represents a fully predictable deterministic pattern
of environment variations.
%The blue curve corresponds to a low transition rate $k_f$.
Here the graph $\lambda(\alpha)$ follows a complex profile
 as $\alpha$ is varied from the region $\alpha<-0.5$, where the transition rates $k_{1,2}(E)$ are always equal to $k_f$,
 to the region $\alpha>0.5$, where there are no transitions between the phenotypes ($k_{1,2}(E)=0$). The average growth rate $\lambda$ has a local peak in the region $\alpha<0$
 and the peak value increases with $k_f$.
 For small transition rates $k_f$, the value
 of $\lambda$ at this local peak is still less than the growth rate $\gamma+0.5\sigma$, which is achieved for $\alpha>0.5$ by the regime without transitions.
 %The local peak value of $\lambda$ increases with increasing $k_f$.
 %At $k_f\approx0.7$, the peak is the same height as the average growth rate for the region $\alpha>0.5$ (the black line in Figure~\ref{fig5}(b)).
 For larger $k_f$, this peak becomes the global maximum, that is the maximum average growth rate is achieved by the bet-hedging (pessimistic) strategy corresponding to an $\alpha<0$. As $k_f$ increases further, the peak shifts towards the point $\alpha=0$.
 %The peak at a negative value of $\alpha$ means that the growth rate is maximized by the predictive switching strategy.
 This behavior agrees with the results of \cite{bac1}. However, we see that for the present model $k_f$ should be large enough
 to favor the bet-hedging strategy; otherwise, the negative effect of the switching cost dominates
 and the strategy forbidding transitions between the phenotypes becomes optimal.

\section*{Discussion}
The model for growth of organisms capable of switching between two stable phenotypes can be viewed as a simple generalization of models that have been previously employed to illustrate the effect of bet-hedging which effectively appears whenever the population either does not switch fast enough \cite{bac1} to follow changes in the environment or whenever the mechanism of following environmental changes is too costly \cite{bac2}.  The generalized model proposed in this paper explicitly accounts for the lag phase often observed after rapid changes in the environment and related phenotype switching, for the continuously and sometimes randomly varying environment as well as for the possibility that switching rates for different phenotypes may change differently with the environmental variations. The model assumes the simplest threshold-like dependence of the phenotype switching rates on the environment that was also used in previously developed models. However, difference in the response of the switching rates to environmental changes for different phenotypes is modeled using different thresholds, with their difference parameterized by variable $\alpha$, for the environmental dependence of the switching rate associated with different phenotypes.

The results presented above demonstrate that there exists a positive threshold difference $\alpha$ at which the growth measured by the Lyapunov's exponent is maximized whenever there is a non-zero lag phase delay and a difference in the growth rates for the two phenotypes.
The positive $\alpha$ decreases the rate of transitions from less to more favored phenotype when favoring is
not strong. Such suppression of back and forth switching agrees with some experimental findings \cite{last, last1}. A slight decrease in the growth rate due to small
fluctuations of the environment from the point where both phenotypes are equally favored can be less dramatic than a drop in the growth rate due to passing through
the lag phase induced by a switching event. This trade off between too much
responsiveness to environmental variations (small $\alpha$), with the associated cost of often transitions, and too much inertia
(large $\alpha$), which leaves too many bacteria in unfavored states, shifts the optimal threshold difference $\alpha$ to the positive range.
In fact, this optimal threshold difference grows with increasing lag phase delay (which is equivalent to a decreasing frequency of transition to the lag phase state).  This type of relationship is suggestive of a resonant behavior where the threshold difference is a tuning parameter that defines a certain characteristic time with respect to the lag phase delay that can be viewed as the cost of switching the phenotype measured in units of time.

The question arises: How can the threshold difference be related to some characteristic time? When the environmental variations are described by a diffusion-like random process, a certain time is needed for the environmental input to pass from one threshold to another if the threshold difference is positive. Larger threshold difference requires larger time for this passage. We suggest that it is this so-called first passage time that is tuned when we vary the threshold difference. The growth maximizing relationship between the first passage time and the lag phase delay time also depends on the effective difference in the growth rates of the two phenotypes.

One can also view the resonance behavior described above from a more traditional stochastic resonance point of view. Consider the situation when the threshold difference is fixed, but the effective strength of the environmental fluctuations is tuned instead. In this case one will observe a growth maximum at a certain strength of the environmental fluctuations. This is similar to a stochastic resonance in bistable systems with a small input having a certain period where signal to noise ratio is maximized at a certain level of noise.

It is worth noting that deterministic environmental input does not lead to the same type of phenomenon, as illustrated by Figure \ref{relay}(b) where the time interval between changes of phenotype switching rates is independent of the threshold difference and is always equal to half the period.
This is interesting because it suggests that threshold difference is only useful in the presence of environmental uncertainty helping the system
to minimize the risk of changing its phenotype switching rate too often.

Emergence of positive threshold difference $\alpha$ can also be viewed as emergence of memory that, in a certain limiting cases, reduces to hysteresis.
The presence of hysteresis becomes apparent if the rate of input variations
slows down, or the input becomes deterministic, but the threshold difference $\alpha$ does not adapt to this change of environmental conditions.
In order to model this situation, we first find the optimal positive $\alpha$ which maximizes the net growth rate of the population
for some stochastic environmental input and then test the system (with the same $\alpha$ and other parameters fixed)
by a modified input.
Since the system response is essentially determined by the ratio of the lag phase delay and the characteristic passage times of the input,
the change of the environmental conditions can be modeled either by slowing down the input variations or, alternatively, by increasing
transition rates between phenotypes.
The presence of hysteresis for positive $\alpha$ is most easily illustrated for the case when $k_u=0$ while $k_f$ and $\delta$ are infinite. The infinitely high transition rates $k_f$ and $\delta$ ensure that the entire population can be treated as if it were in a favored (faster growing) phenotype when the environmental input is outside the interval $0.5-\alpha\le E\le 0.5+\alpha$.
 Furthermore, as $k_u=0$, there are no transitions between phenotypes when the environmental input is inside of this interval. In this case the interval $0.5-\alpha\le E\le 0.5+\alpha$
 can be viewed as the true bi-stability interval.  Therefore, all the organisms in the population retain their current phenotype until the environment exits the bi-stability interval. This is exactly the behavior described by the bi-stable relay illustrated in Figure \ref{relay}, which represents the simplest form of hysteretic memory. While hysteretic memory holds in the case considered above for all possible rates of environmental input variations, it holds approximately even if $k_f$ and $\delta$ are finite and $k_u$ is non-zero as long as the ratios $k_f/k_u$, $\delta/k_u$ are large and the environmental input varies sufficiently slowly. %(but not too slowly). }

Thus, the model suggests that hysteresis found in laboratory experiments
is a manifestation of more complex memory of multi-phenotype biological systems
and that hysteresis becomes apparent because environmental conditions used in experiments
are less variable or more predictable than natural environments.
Moreover, the stochastic resonance mechanism can be effective if the magnitude of hysteresis is tuned to the strength of natural environmental
fluctuations in such a way that ensures correlation of the corresponding passage times with the lag phase delay.
This suggests that measurements of hysteresis and lag phase can help characterize fluctuations of the natural environment.

{ Finally, it is reasonable to interpret the model as one where hysteretic memory is attributed to individual organisms rather than the entire population since no interaction between organisms has been explicitly included in the model and the lag phase delay inducing the observed behavior is the property of the individual organisms.}

 %\textcolor{red}{Gena, should we have here a paragraph discussing that our model produces
 %a bet-hedging mechanism when $\alpha$ is negative, hence excluding hysteresis in this case
 %- something like the following paragraph (modification is needed) - or would such a paragraph just distract the reader?
 %``{On the other hand, if $\alpha$ is negative, while $k_u=0$ while $k$ is infinite, both states are out of favor and cells will switch into the holding states $w$ and $z$ as soon as the %environmental signal enters the bi-stability interval. Should any cell switch to one of the growing phenotypes, it will immediately switch back to a holding (non-growing state). Thus, no %hysteresis would occur for negative α. Instead, the population will cycle between two different holding states in this case as long as the environmental variable remains in the %bi-stability interval.}''
 %}

% You may title this section "Methods" or "Models".
% "Models" is not a valid title for PLoS ONE authors. However, PLoS ONE
% authors may use "Analysis"
%\section*{Materials and Methods}

% Do NOT remove this, even if you are not including acknowledgments
%\section*{Acknowledgments}

%DR acknowledges the support of the Russian Foundation for basic Research through grant 10-01-93112
%and the Alexander von Humboldt Foundation (Germany).

%\section*{References}
% The bibtex filename
\bibliography{template}

\end{document}